\documentclass[11pt, a4paper, onecolumn]{article}

\usepackage{fullpage}
\usepackage{graphicx}
\usepackage[english]{babel}
\usepackage[center]{caption}
\usepackage{mathtools}
\usepackage{mathptmx}    
\usepackage{latexsym}           
\usepackage{url}
\usepackage{multirow}
\usepackage{amssymb}
\usepackage{amsthm}
\usepackage{amssymb}
\usepackage{amsmath}
\usepackage{lineno}
\usepackage{setspace}
\usepackage{hyperref}

\newcommand{\Keywords}[1]{\par\noindent {\bf {\em Keywords}: }#1 }

\begin{document}

\title{On Projection Based Operators in $l_p$ space for Exact Similarity Search}

\author{Andreas Wichert\\ \small \texttt{andreas.wichert@ist.utl.pt}\\
\and
Catarina Moreira\\ \small \texttt{catarina.p.moreira@ist.utl.pt}
\and
\\ \small Instituto Superior T\'{e}cnico, INESC-ID\\ \small Av. Professor Cavaco Silva, 2744-016 Porto Salvo, Portugal\\  
\\ \small The original publication will be available at: \\ \small Fundamenta Informaticae: Annales Societatis Mathematicae Polonae, 136: 1-14, 2015\\
 \small \text{\url{http://iospress.metapress.com/content/97l2163185243216/}}
}

\date{}

\maketitle

\doublespace

\begin{abstract}

We investigate exact indexing for high dimensional $l_p$ norms based on the 1-Lipschitz property and projection operators. 
The orthogonal projection that satisfies the 1-Lipschitz property for the $l_p$ norm is described. The adaptive projection defined by the first principal component is introduced. 
\end{abstract}


\section{Introduction} 

While there are relatively efficient approximate similarity search algorithms, it is widely supposed that the exact search suffers from dimensionality  \cite{Pestov2012}. Thus, solving the problem in the most general case for an arbitrary dataset is impossible. 
We investigate exact indexing for a vector space $V$ and a distance function $d$.  Exact indexing is based on exact similarity search, and no data points are lost during range queries. For a range query vector $\textbf{y}$ from a collection of $s$ vectors,
\[ \textbf{x}_1,  \textbf{x}_2,  \textbf{x}_3, \cdots,  \textbf{x}_s \] 
\textit{all} vectors $\textbf{x}_i$ that are $\epsilon$-similar according to the distance function $d$ are searched
  \begin{equation}
 d( \textbf{x}_i,\textbf{y} ) <  \epsilon.
\end{equation}
In approximate indexing, the data points that may be lost as some distances are distorted. Approximate indexing  \cite{Indyk98},  \cite{Indyk04} seems to be in some sense free from the curse of dimensionality, \cite{Pestov2012}.
 Distance-based exact indexing is based on the 1-Lipschitz property \cite{Pestov2012}.  A mapping function $F()$ maps two vectors $\bf{x}$ and $\bf{y}$ into a lower dimensional space, where $d$ is a metric in the original space and $d_{feature}$ is a metric in the feature space that satisfies the 1-Lipschitz property
\begin{equation}
d_{feature}(F( \textbf{x} ),F( \textbf{y} )) \leq d( \textbf{x},\textbf{y} ).
\end{equation}
This equation is also known as the lower bounding postulate \cite{Faloutsos94b}, \cite{Faloutsos99}. Using the 1-Lipschitz property, a bound that is valid in both spaces can be determined. The distance from similar vectors to a query vector $\textbf{y}$ is smaller or equal in the original space and, consequently, is smaller or equal in the lower dimensional space as well. 
During the computation, all the points below the bound are discarded. In the second step, the wrong candidates are filtered by comparisons in the original space.
The application of the 1-Lipschitz property as used in metric trees and pivot tables does not resolve the curse of dimensionality, as shown in \cite {Pestov2011}. For high-dimensional spaces, the functions that obey the 1-Lipschitz property discard fewer points as the number of dimensions grows \cite {Pestov2011}. The number of points discarded drops as fast as the number of dimensions grows. As stated in \cite{Pestov2012}, every 1-Lipschitz function concentrates sharply near its mean (or median) value, which results from the fact that a sphere with a constant radius increases exponentially with growing dimensions. A linear radial increase leads to an exponential increase of points inside the sphere \cite{Bohm01}, \cite{Pestov2012}, which leads to a degradation of the method's performance. This situation leads to the ``curse of dimensionality'', which states that for an exact nearest neighbor, any algorithm for high dimension $d$ and $n $ objects must either use an $n^d$-dimension space or have a query time of $n \times d $ \cite{Bohm01}, \cite{Pestov2012}.

However, \cite{Wichert08}, \cite{Wichert10}, and \cite{Wichert12} show how the recursive application of the 1-Lipschitz property can be used to overcome the curse of dimensionality for certain cases of points equally distributed by subspace trees. A high-dimensional space is divided into low-dimensional sub-spaces \cite{Wichert08}, \cite{Wichert10}. In the low-dimensional sub-spaces, 1-Lipschitz functions can be successfully applied. 
The main contributions of this paper are as follows:
\begin{itemize}
\item Introduction of a new adaptive projection. The optimal projection is not fixed by orthogonal projection but learned.
\item Extension of the technique beyond the Euclidean norm ($l_2$). Many applications rely on the $l_1$ norm. It is shown that $l_1$ norm gives better results than $l_2$ norm.
\item Simplification of the mathematical framework.
\end{itemize}
The paper is organized as follows: 
\begin{itemize}
\item We review the projection operators.
\item We introduce the adaptive projection and the $l_p$ norm dependency.
\item The adaptive projection and the $l_p$ norm dependency are integrated into the subspace tree.
\item We empirically compare the adaptive projection with the orthogonal mapping. We empirically compare the $l_1$, $l_2$, $l_4$ and $l_{\infty}$ norms.
\end{itemize}

\section{Projection Operators}

Ideally, the mapping function $F()$ should preserve the exact distances \cite{Faloutsos94b}, \cite{Faloutsos99}.  An example of such a function for real vectors is a norm preserving linear operator $Q$. Such an operator can be represented by an orthogonal matrix with $Q^T=Q^{-1}$ performing a rotation or a reflection. An example of such an operator is the Karhunen-Lo\`{e}ve transform, which rotates the coordinate system in such a way that the new covariance matrix will be diagonal, resulting in each dimension being uncorrelated.
A mapping that reduces the dimensionality of a vector space can be represented by a projection operator in a Hilbert space, which extends the two or three dimensional Euclidean space to spaces with any finite or infinite number of dimensions.  In such a space, the Euclidean norm is induced by the inner product
\begin{equation}
\|\textbf{x}\|=\sqrt{ \langle  \textbf{x}|\textbf{x} \rangle}. 
\end{equation}
If $W$ is a subspace of $V,$ then the orthogonal complement of $W$ is also a subspace of $V.$ The orthogonal complement $W^\bot$ is the set of vectors
\begin{equation}
W^\bot=\{ \textbf{y} \in V |  \langle  \textbf{y}|\textbf{x} \rangle=0~~  \textbf{x} \in V \} 
\end{equation}
and
\begin{equation}
V=W \oplus W^\bot. 
\end{equation}
Each vector $\textbf{x} \in V$ can be represented as $\textbf{x}= \textbf{x}_W + \textbf{x}_{W^\bot} $ with $\textbf{x}_W \in W$ and $ \textbf{x}_{W^\bot} \in W^\bot$. The mapping $P \cdot \textbf{x}=\textbf{x}_W$ is an orthogonal projection. Such a projection is always a linear transformation and can be represented by a projection matrix $P$. The matrix is self-adjoint with $P=P^2$.  An orthogonal projection can never increase a norm
\begin{equation}
 \|P \cdot \textbf{x}\|^2 = \|\textbf{x}_W\|^2 \leq \| \textbf{x}_W  \|^2+ \| \textbf{x}_{W^\bot} \|^2 =\| \textbf{x}_W + \textbf{x}_{W^\bot} \|^2 = \|\textbf{x}\|^2.
 \end{equation}
Using  the triangle inequality
\begin{equation}
\| \textbf{x} + \textbf{y} \| \leq \| \textbf{x} \| + \|\textbf{y}\| 
\end{equation}
setting
\begin{equation}
\|  \textbf{x} \| =  \| \textbf{y} + (\textbf{x} -  \textbf{y}) \| \leq  \|\textbf{y}\|+\|\textbf{x}-\textbf{y}\| 
\end{equation}
the tighten triangle inequality 
\begin{equation}
 \|  \textbf{x} \| -  \| \textbf{y} \|  \leq \|\textbf{x}-\textbf{y} \|. 
\end{equation}
follows.
From the fact that the orthogonal projection can never increase the norm and the tightened triangle inequality, any orthogonal projection operator has the 1-Lipschitz property
 \begin{equation}
 \| P \cdot \textbf{x} \| -  \| P \cdot \textbf{y} \| | \leq \| P \cdot\textbf{x}- P \cdot \textbf{y}\| = \|  P \cdot( \textbf{x}- \textbf{y}) \|  \leq \|\textbf{x}-\textbf{y} \|.  
\end{equation}
It follows that any projection satisfies the 1-Lipschitz property, which means that the lower bounding postulate \cite{Faloutsos94b},\cite{Faloutsos99} and any orthogonal projection are satisfied. For example, the ``Quadratic Distance Bounding'' theorem is satisfied \cite{Faloutsos94b}.   There is no the need for a more complicated proof based upon the unconstrained minimization problem using Lagrange multipliers \cite{Faloutsos94b}.   

\subsection{Projection onto one-dimensional subspace}

For $\| \textbf{p} \|=1$, $\textbf{p} \cdot \textbf{p}^\top$ is an orthogonal projection onto a one-dimensional space generated by $ \textbf{p} $. For example for the vector 
 \begin{equation}
 \textbf{p}= \left( \frac{1}{\sqrt{n}}, \frac{1}{\sqrt{n}}, \cdots, \frac{1}{\sqrt{n}} \right) 
 \end{equation}
the orthogonal projection from $R^n$ onto one-dimensional space $R$ is
 \begin{equation}
 P=\textbf{p} \cdot \textbf{p}^\top= \left(  \begin{array}{cccc}
\frac{1}{n} & \frac{1}{n}  & \cdots &  \frac{1}{n} \\
\frac{1}{n} & \frac{1}{n}  & \cdots &  \frac{1}{n} \\
\vdots & \vdots & \ddots & \vdots \\
\frac{1}{n} & \frac{1}{n}  & \cdots &  \frac{1}{n} \\
\end{array} \right).
  \end{equation}
 For $f \leq n$  orthogonal subspaces 
  \begin{equation}
 R^n=E_1 \oplus E_2   \oplus \ldots  \oplus E_f 
 \label{decomp}
 \end{equation} 
 of the vector space $R^n$  we can define a projection $P: R^n \mapsto R^f$ as a sum of $f$ projections onto one dimensional space
  \begin{equation}
  P= \textbf{p}_1\cdot \textbf{p}_1^\top +\textbf{p}_2 \cdot \textbf{p}_2^\top  \ldots  +\textbf{p}_f \cdot \textbf{p}_f^\top 
  \label{Proj_eq}
  \end{equation} 
  with $ \textbf{p}_i \cdot \textbf{p}_i^\top :  E_i \mapsto R$ and $\| \textbf{p}_i \|=1$.  
  The  1-Lipschitz property  of the projection from the subspace $E_i$  the one dimensional space $R$  is
 \begin{equation}
 \left| \| \textbf{p}_i \cdot \textbf{p}_i^\top   \cdot \textbf{x} \| -  \|  \textbf{p}_i \cdot \textbf{p}_i ^\top\cdot \textbf{y} \| \right| \ \leq 
  \left| \|\textbf{x} \| -\| \textbf{y}\| \right|  \leq  
  \|\textbf{x}-\textbf{y}\|. 
  \label{1Lip_eq}
   \end{equation}
 The projection $P$, represented by Equation \ref{Proj_eq}, should distort the distances between the vector space $R^n$ and $R^f$ as little as possible. As a consequence, the distortion for each subspace $E_i$ should be minimized. Because of the 1-Lipschitz property for the one-dimensional space, according to the Equation \ref {1Lip_eq}, we need to minimize the distance in the one-dimensional space between the length of the vector and the length of its projected counterpart
 \begin{equation}
\left|  \| \textbf{p}_i \cdot \textbf{p}_i^\top   \cdot \textbf{x} \|  -\|\textbf{x}  \| \right|.
 \end{equation}  
Suppose the dimensionality of the subspace $E_i$ is $m$. We define the vector  $\textbf{a}$ as 
 \begin{equation}
  \textbf{a}    = \textbf{p}_i\cdot \textbf{p}_i^\top    \cdot \textbf{x}. 
 \end{equation} 
It follows that 
  \begin{equation}
 a=   \sqrt{m} \cdot \alpha =  \|  \textbf{a} \|   =\| \textbf{p}_i\cdot \textbf{p}_i^\top    \cdot \textbf{x} \|
 \end{equation} 
and with
 \[ a_1=a_2=...=a_k=...=a_m=\alpha \]
\begin{equation}
\textbf{a}=(a_1,a_2,..a_k,..,a_m).
\end{equation} 
 With $a$ being the length of the projected vector we preform the following operation
\begin{equation}
 \min \{ |a  -\|\textbf{x}  \|| \}.  
 \end{equation}
 From the tighten triangle inequality, it follows that
\begin{equation}
\min \{ a  -\|\textbf{x}  \| \} \leq   \min \{  \| \textbf{a} - \textbf{x}  \| \} 
 \end{equation}
 according to the Euclidean distance function. To minimize the Euclidean metric $ \| \textbf{a} - \textbf{x} \|$, how do we choose the value of $\alpha$ \cite{Wichert12}? It follows that 
\begin{equation}
 \min_\alpha \left( \sqrt{( x_1-  \alpha)^2 +(x_2-  \alpha)^2 +...+(x_m-  \alpha)^2 } \right)  
 \end{equation}
\begin{equation}
 0=\frac { \partial d (\vec{x},\vec{a} ) } { \partial \alpha } =  \frac { m \cdot \alpha -  \left( \sum_{i=1}^{m} x_i  \right) } { \sqrt { m \cdot \alpha^2 + \sum_{i=1}^{m} x_i^2-2 \cdot  \alpha \cdot \left(  \sum_{i=1}^{m} x_i   \right)}} 
 \end{equation}
with the solution 
\begin{equation}
 \alpha = \frac{\sum_{i=1}^{m} x_i}{m}  
  \end{equation}
which is the mean value of the vector $\textbf{x}$.  
It follows
\begin{equation}
 a=   \sqrt{m} \cdot \alpha = \sqrt{m} \cdot  \frac{\sum_{i=1}^{m} x_i}{m}    =\| \textbf{p}_i \cdot \textbf{p}_i^\top    \cdot \textbf{x}  \| 
 \end{equation}
with the corresponding projection matrix $P_i$
\begin{equation}
 P_i=\textbf{p}_i \cdot \textbf{p}_i^\top = \left(  \begin{array}{cccc}
\frac{1}{m} & \frac{1}{m}  & \cdots &  \frac{1}{m} \\
\frac{1}{m} & \frac{1}{m}  & \cdots &  \frac{1}{m} \\
\vdots & \vdots & \ddots & \vdots \\
\frac{1}{m} & \frac{1}{m}  & \cdots &  \frac{1}{m} \\
\end{array} \right).
  \end{equation}
$P_i$ is generated by the normalised vector $\textbf{p}_i$ 
\begin{equation}
 \textbf{p}_i =  \left( \frac{1}{\sqrt{m}}, \frac{1}{\sqrt{m}}, \cdots, \frac{1}{\sqrt{m}} \right).
 \end{equation}
 which indicates the direction of the $m$-secting line, which is a continuous map from a one-dimensional space to an $m$-dimensional space given by 
\begin{equation}  
\begin{array}{c}
 x_1=x_1\\
 x_2=x_1\\ 
 x_3=x_1\\
\vdots\\ 
 x_m=x_1\\
 \end{array}
 .
  \end{equation}  
For $m=2$, this equation is the bisecting line with $x_1=x_2$ or, represented as a curve, 
\begin{equation}  
\begin{array}{c}
 x_1=x_1\\
 x_2=x_1\\ 
 \end{array}
  \end{equation}  
which, for uncorrelated data $P_i$, is the best projection onto one dimension, as indicated in next section. The projection can be computed efficiently without needing matrix operations as the mean value of the vector multiplied with the square root of its dimensionality.\begin{equation}   
 \sqrt{m} \cdot \ \frac{\sum_{i=1}^{m} x_i}{m}
= \left|  \left|  \left( \begin{array}{c}
\frac{\sum_{i=1}^{m} x_i}{m}  \\
\frac{\sum_{i=1}^{m} x_i}{m}  \\
\vdots\\
\frac{\sum_{i=1}^{m} x_i}{m}  \\
\end{array}
 \right) \right| \right|
= \left|  \left|  \left(  \begin{array}{cccc}
\frac{1}{m} & \frac{1}{m}  & \cdots &  \frac{1}{m} \\
\frac{1}{m} & \frac{1}{m}  & \cdots &  \frac{1}{m} \\
\vdots & \vdots & \ddots & \vdots \\
\frac{1}{m} & \frac{1}{m}  & \cdots &  \frac{1}{m} \\
\end{array}
 \right)
\cdot
\left(  \begin{array}{c}
x_1\\
x_2\\
\vdots\\
x_m\\
\end{array}
 \right) \right| \right|.
  \end{equation}
A projection $P: R^n \mapsto R^f$  
given the decomposition into $f$ orthogonal spaces according to Equation \ref{decomp}
is composed of a sum of $f$ projections onto a one-dimensional space.
Each projection is a projection on an $m$-secting line with $P_i:R ^{m} \mapsto R$. 
The method works with the space split in any way. For simplicity, we assume that the $n$-dimensional space is split into $f$ equal-dimensional subspaces.  In this case, the projections are efficiently computed as the mean value of each sub-vector. The corresponding mean values are multiplied with the constant $c=\sqrt{m}=\sqrt{\frac{n}{f}}$.  The selection of the division can be determined by empirical experiments in which we relate $m$ to $n$ with the constraint that $n$ is divisible by $m$.  

\subsection{The First Principal Component}

The covariance matrix represents the tendency of two dimensions varying in the same direction as indicated by the data points.  The Karhunen-Lo\`{e}ve transform rotates the coordinate system in such a way that the new covariance matrix will be diagonal. Therefore, each dimension will be uncorrelated. The transformation is described by an orthonormal matrix, which is composed of the normalized eigenvectors of the covariance matrix.  The squares of the eigenvalues represent the variances along the eigenvectors. The first principal component corresponds to the normalized eigenvector $\textbf{z}$ with the highest variance.

$\| \textbf{z} \|=1$ with $Z=\textbf{z} \cdot \textbf{z}^\top$ is the best projection onto one-dimensional space because, in a Hillbert space, the first principal component passes through the mean and minimizes the sum of squares of the distances of the points from the line. It follows that
\begin{equation}
   \| \textbf{x} \|  \geq \| P \cdot  \textbf{x} \|  \geq   \| Z \cdot  \textbf{x} \|. 
\end{equation}
For uncorrelated data, $Z=P$ represents the projection on the $m$-secting line. For correlated data contrary to the projection on the $m$-secting line, all the components of the vector $\textbf{z}$ do not need to be equal, and the projection cannot be computed efficiently.
For a vector $\textbf{o}$ of length $\sqrt{m}$ in the direction of the $m$-secting line where $P$ is the projection on the $m$-secting line, 
 \begin{equation}
\textbf{o}=\underbrace{(1,1,1,\cdots,1)}_{m}
 \end{equation}
it follows that
 \begin{equation}
    \sqrt{m}= \| \textbf{o} \|= \| P \cdot  \textbf{o} \| \geq   \| Z \cdot  \textbf{o} \|  \geq 1. 
  \end{equation}  
 The value 
  \begin{equation}
 \sqrt{m} - \| Z \cdot  \textbf{o} \|
 \label{eq:abstand}
 \end{equation}
 indicates the diversion from the $m$-secting line with value $0$ corresponding  to uncorrelated data and  $\sqrt{m}-1$ one dimension data with 
 \begin{equation}
  \sqrt{m}-1  \geq  \sqrt{m} - \| Z \cdot  \textbf{o} \|  \geq 0.    
  \end{equation}
For a given decomposition into $f$ orthogonal spaces according to the Equation \ref{decomp} the data points are mapped into corresponding subspaces $E_i$. For each subspace $E_i$ the covariance matrix $C_i$ is computed. In the next step for each  covariance matrix $C_i$  the first principal component with the highest variance is determined. It is represented it by the normalised eigenvector  $\textbf{z}_i$. Each projection 
 \begin{equation}
Z_i=\textbf{z}_i \cdot \textbf{z}_i^\top
 \end{equation}
 is a projection onto the first principal component with $Z_i:R ^{m}  \mapsto R$.  An adaptive projection $A: R^n \mapsto R^f$, given the decomposition into $f$ orthogonal spaces according to the Equation \ref{decomp}, is composed of a sum of $f$ projections $Z_i$ onto a one-dimensional space.
\begin{equation}
 A= \textbf{z}_1 \cdot \textbf{z}_1^\top +\textbf{z}_2 \cdot \textbf{z}_2^\top  \ldots  +\textbf{z}_f \cdot \textbf{z}_f^\top. 
 \end{equation}
The method works under any splitting of the space, such as the projection $P: R^n \mapsto R^f$. 
 
\subsection{$l_p$ norm dependency}

Some applications require distance functions that differ from the Euclidian distance function. In addition to the Euclidean distance function, the Manhattan distance and the Chebyshev distance function are commonly used. In the following, we generalize the Euclidean norm to the $l_p$ norm that induces a corresponding metric. The $l_p$ norm is defined as the following (for $p=2$ it is the Euclidean norm): 
 \begin{equation}
 \|\textbf{x}\|_p=\left( |x_1|^p+|x_2|^p+\cdots+|x_m|^p \right)^{\frac{1}{p}} 
 \end{equation}
$l_p$ norms are equivalent and the following relation holds for $0 < q < p$
 \begin{equation}
  \|x\|_p \leq \|x\|_q  \leq  m^{ \frac{1}{q}-\frac{1}{p} } \cdot  \|x\|_p 
   \end{equation}
and
 \begin{equation}
 m^{ \frac{1}{p}-\frac{1}{q} } \cdot \|x\|_q \leq \|x\|_p  \leq    \|x\|_q. 
  \end{equation}
The tighten triangle inequality is valid in any $l_p$ norm due to the definition of norm. Because the $l_p$ norms are equivalent  the following equation is valid as well for any $l_p$ norm
 \begin{equation}
   \| P \cdot \textbf{x} \|_p -  \| P \cdot \textbf{y} \| |_p \leq \| P \cdot\textbf{x}- P \cdot \textbf{y}\|_p = \|  P \cdot( \textbf{x}- \textbf{y}) \|_p  \leq \|\textbf{x}-\textbf{y}\|_p  
   \end{equation} 
and
 \begin{equation}
  \| Z \cdot  \textbf{x} \|_p   \leq \| P \cdot  \textbf{x} \|_p  \leq  \| \textbf{x} \|_p.  
  \end{equation}
The  linear projection operator $P$ has  the 1-Lipschitz property in any $l_p$ norm and
 \begin{equation} 
 m^\frac{1}{p} \cdot \ \frac{\sum_{i=1}^{m} x_i}{m}
= \left|  \left|  \left( \begin{array}{c}
\frac{\sum_{i=1}^{m} x_i}{m}  \\
\frac{\sum_{i=1}^{m} x_i}{m}  \\
\vdots\\
\frac{\sum_{i=1}^{m} x_i}{m}  \\
\end{array}
 \right) \right| \right|_p
= \left|  \left|  \left(  \begin{array}{cccc}
\frac{1}{m} & \frac{1}{m}  & \cdots &  \frac{1}{m} \\
\frac{1}{m} & \frac{1}{m}  & \cdots &  \frac{1}{m} \\
\vdots & \vdots & \ddots & \vdots \\
\frac{1}{m} & \frac{1}{m}  & \cdots &  \frac{1}{m} \\
\end{array}
 \right)
\cdot
\left(  \begin{array}{c}
x_1\\
x_2\\
\vdots\\
x_m\\
\end{array}
 \right) \right| \right|_p.
  \end{equation}
The projection $P$ can be computed efficiently without needing a matrix operation as the mean value of the vector multiplied with the constant $c= m^\frac{1}{p}$. For the dimension $m$, for the $l_1$ norm, $c=m$, for the $l_2$ norm, $c= \sqrt{m}$, and for the $l_\infty$ norm, $c= 1$. A lower $l_p$ norm corresponds to a higher  constant $m \geq c \geq 1$ and less information loss. We cannot gain any advantage of the 1-Lipschitz property using the different $l_p$ norms.  The behavior of the constant $c$ is related to the equivalence of the norms relation. For example, the $l_1$ and $l_2$ relation is 
 \begin{equation}
 \|x\|_2 \leq \|x\|_1  \leq    \sqrt{m} \cdot \|x\|_2. 
 \end{equation}
For   $\| \textbf{q} \|_p=1$ with $Q=\textbf{q}^\top \cdot \textbf{q}$ is a mapping onto one dimensional space generated by $ \textbf{q} $. It is not a projection for $p > 2$ because the matrix is not self-adjoint with $Q=Q^2$. The mapping on the $m$-secting line. the operator can be understood as 
 \begin{equation}
\ Q=\textbf{q}^\top \cdot \textbf{q} = \left(  \begin{array}{cccc}
\frac{1}{m^\frac{2}{p}} & \frac{1}{m^\frac{2}{p}}  & \cdots &  \frac{1}{m^\frac{2}{p}}\\
\frac{1}{m^\frac{2}{p}} & \frac{1}{m^\frac{2}{p}}  & \cdots &  \frac{1}{m^\frac{2}{p}}\\
\vdots & \vdots & \ddots & \vdots \\
\frac{1}{m^\frac{2}{p}} & \frac{1}{m^\frac{2}{p}}  & \cdots &  \frac{1}{m^\frac{2}{p}}\\
\end{array} \right).
 \end{equation}
$Q$ is generated by the $l_p$ normalized vector $\textbf{q}$ indicating the direction of the $m$-secting line.
 \begin{equation}
 \textbf{q}^\top =  \left( \frac{1}{m^\frac{1}{p}}, \frac{1}{m^\frac{1}{p}}, \cdots, \frac{1}{m^\frac{1}{p}} \right). 
 \end{equation}
The mapping can be computed efficiently without requiring matrix operations as the mean value of the vector multiplied with the constant 
$d=m^\frac{p-1}{p}$.  
 \begin{equation} 
 m^\frac{p-1}{p} \cdot \ \frac{\sum_{i=1}^{m} x_i}{m}
= \left|  \left|  \left( \begin{array}{c}
\frac{\sum_{i=1}^{m} x_i}{m^\frac{2}{p}}  \\
\frac{\sum_{i=1}^{m} x_i}{m^\frac{2}{p}}  \\
\vdots\\
\frac{\sum_{i=1}^{m} x_i}{m^\frac{2}{p}}  \\
\end{array}
 \right) \right| \right|_p
= \left|  \left|  \left(  \begin{array}{cccc}
\frac{1}{m^\frac{2}{p}} & \frac{1}{m^\frac{2}{p}}  & \cdots &  \frac{1}{m^\frac{2}{p}}\\
\frac{1}{m^\frac{2}{p}} & \frac{1}{m^\frac{2}{p}}  & \cdots &  \frac{1}{m^\frac{2}{p}}\\
\vdots & \vdots & \ddots & \vdots \\
\frac{1}{m^\frac{2}{p}} & \frac{1}{m^\frac{2}{p}}  & \cdots &  \frac{1}{m^\frac{2}{p}}\\
\end{array}
 \right)
\cdot
\left(  \begin{array}{c}
x_1\\
x_2\\
\vdots\\
x_m\\
\end{array}
 \right) \right| \right|_p.
 \end{equation}
However this mapping can  increase a norm. For the norm $l_p$ the induced matrix norm is 
 \begin{equation} 
 \| Q \|_p= \max_{\|x\_p|} \|Q \cdot \textbf{x}\|_p
 \end{equation}
and for $\textbf{x}=\textbf{q}$ 
  \begin{equation} 
  \|Q\|_p= m^\frac{p-2}{p}. 
  \end{equation}
 It follows that for $p >2$
 \begin{equation} 
 \|Q \cdot \textbf{x}\|_p > \| \textbf{x}\|_p
 \end{equation}
the norm is increased.
Only for $p  \leq 2$ the norm is not increased with $l_2$ the projection $P$ and $l_1$ the simple mean value. 

\section{Subspace tree revisited}

An adaptive projection, $A: R^n \mapsto R^f$, maps two vectors, $\bf{x}$ and $\bf{y}$, into a lower-dimensional space and satisfies the 1-Lipschitz property:
\begin{equation} 
 \| A \cdot \textbf{x} - A \cdot \textbf{y} \| |_p \leq \ \|\textbf{x}-\textbf{y}\|_p.  
 \end{equation}
 Using the 1-Lipschitz property, a bound that is valid in both spaces can be determined. The distance of similar vectors to a query vector $\textbf{y}$ is smaller or equal in the original space of dimensionality $n$, and, consequently, it is also smaller or equal in the lower-dimensional space of the dimensionality $f$.  During the computation, all the points below the bound are discarded. In the second step, the wrong candidates are filtered by comparisons in the original space. The number of points discarded drops as fast as the relation between the dimensionalities $\frac{n}{f}$ grows. Depending on the correlation between the dimensionalities, the 1-Lipschitz property is only useful if the relation is sufficiently small with 
\begin{equation} 
 \frac{n}{f} \leq d
 \end{equation}
where $d$ varies between $2 \leq d \leq 16$ in relation to the data set. However, high-dimensional indexing requires that the mapping $F: R^n \mapsto R^d$ with $ n\gg d$ satisfies the 1-Lipschitz property. For such a function, only a tiny fraction of the points of a given set are below the bound. Thus, the majority of the points have to be filtered by comparisons in the original space. Therefore, no speed up, compared tp the  use of a simple list matching, can be achieved, as proclaimed by the conjecture ``the curse of dimensionality''. 
If at least some points of a given set are below the bound, there is a way to build a recursive function that achieves a considerable speed up using a simple list matching. Motivated by the divide and conquer principle and the tree structure, one can build such a function recursively, indicating that the ``the curse of dimensionality'' conjecture is \textit{wrong} for some data sets. It is well known that, for a dimensionality $d$ ($2 \leq d \leq 16$), metric index trees operate efficiently. Thus, in the next step we define an efficient indexing structure that builds on the mapping $F: R^n \mapsto R^d$ that satisfies the 1-Lipschitz property, with $F$ being a projection or an adaptive projection.

Suppose there exist a sequence of subspaces $U_0, U_1, U_2, \ldots, U_t$ with $R^n=U_0$ and $R^d=U_t$ in which each subspace is a subspace of another space
\begin{equation} 
U_0 \supset U_1 \supset U_2 \supset \ldots   \supset U_t  
 \end{equation}
and with $dim(U_i)$ indicating the dimension of the subspace $U_i$
\[dim(U_0) > dim(U_1) > dim(U_2) \ldots >  dim(U_t) \] and the relation between neighbouring subspaces is sufficiently small  with
\begin{equation} 
 \frac{dim(U_0)}{dim(U_1)} \leq d, \frac{dim(U_1)}{dim(U_2)} \leq d, \ldots \frac{dim(U_{t-1})}{dim(U_t)} \leq d. 
  \end{equation}
We define a family of projections for the  sequence of subspaces (either adaptive  or not) with the following 
\begin{equation} 
 A_1: U_0 \mapsto U_1; A_2: U_1 \mapsto U_2;  \ldots ; A_t: U_{t-1} \mapsto U_t. 
  \end{equation}
The family of projections defines the sequence of subspaces.
Given a bound $\epsilon$ and a query vector $\textbf{y}$ for each subspace including the original space $U_0$, certain points are below the bound. For each subspace $U_i$, the number of points below the bound $\epsilon$ is indicated by the value $\sigma_i$.
It follows that
\begin{equation} 
 \sigma_0 <  \sigma_1 <  \ldots  < \sigma_t < s 
  \end{equation}
where $s$ is the size of the data set. The resulting computing cost given a bound $\epsilon$ and a query vector $\textbf{y}$ is
\begin{equation}
cost_s=\sum_{i=1}^{t} \sigma_i \cdot dim(U_{i-1}) + s  \cdot dim(U_t).
\label{eq:lsubcost} 
\end{equation} 
The cost of list matching is
 \begin{equation} 
cost_l=s  \cdot dim(U_0)
 \end{equation}
The saving $cost_s < cost_l $ is related to the bound $\epsilon$. Empirical experiments suggest that $cost_s \ll cost_l$ for a bound with $\sigma_0 < d$.  

The described projection based method cannot be applied to sparse representation, as present in the vector space model  \cite{Yates99}. 

\subsection{Tree isomorphy }

The isomorphy to a tree results from the assumption that the value of $\sigma_i$ is reciprocal to the preceding dimensionality \cite{Wichert10}.  Therefore, a bigger dimensionality $dim(U_{i+1})$ results in a smaller value $\sigma_i$, and vice versa. We can express this relation by 

 \begin{equation}
 const \cdot \sigma_i  =\frac{1}{dim(U_{i+1})}.
 \label{eq:rep0} 
 \end{equation}
The value of $const$ is dependent on the data set and its norm.
The value of $\sigma_i$ is reciprocal to the preceding dimensionality $dim(U_{i+1})$ (Equation \ref{eq:rep0}), and the computing costs are expressed by
 \begin{equation}
cost_s \approx 1/const \cdot \left( \frac{dim(U_0)}{dim(U_1)}+\frac{dim(U_1)}{dim(U_2)}+\ldots+\frac{dim(U_{n-1})}{dim(U_t)} \right)+ dim(U_t) \cdot s. 
\label{eq:hiecost} 
\end{equation}
Supposing $d= dim(U_t)$ and  $n= dim(U_0)$
 \begin{equation}
cost_s \approx 1/const \cdot   d \cdot  \log_d(n-d)   + d  \cdot s. 
\label{eq:hiecost2} 
\end{equation}
For a dimension $d$, the metric index trees operate efficiently with a nearly logarithmic search time. 
For the bound with $\sigma_0 < d$, the value $1/const \ll s$
 \begin{equation}
cost_s \approx 1/const \cdot   d \cdot  \left( \log_d(n)-1\right)   + d \cdot \log_d(s). 
\label{eq:hiecost3} 
\end{equation}
It follows that the lower bound of the computational cost is 
\begin{equation}
 \Omega(\log(n) + \log( s)). 
 \end{equation}

\section{Examples of  $\epsilon$  similarity}

The $\epsilon$ range queries depends on the adequate value of $\epsilon$. A method for the estimation of such a value is described in \cite{Wichert08}. 
Let $DB$ be a database of $s$ multimedia objects $\textbf{x}^{(i)}$ represented by vectors of dimensionality $n$ in which the index $i$ is an explicit key identifying each object
\begin{equation}
\{\textbf{x}^{(i)} \in DB | i \in \{1.. s\}\}. 
 \end{equation}
For a query object  $\textbf{y}$, $all$ objects $\textbf{x}^{(i)}$  are searched that are $\epsilon$-similar 

\begin{equation}
 \| \textbf{x}^{(i)} - \textbf{y} \|_p < \epsilon.
  \end{equation}
For high dimensional vector space such a set can be determined  by a  list matching over the whole data set.

\subsection{Computational procedure}

For the database $DB$ that is projected into the subspace $U_k$,
\begin{equation}
 \{U_k(\textbf{x})^{(i)} \in U_k(DB) | i \in \{1.. s\}\}. 
  \end{equation}
The algorithm to determine all $\epsilon$-similar objects is composed of two loops. The first loop iterates over the elements of the database $DB$, and the second iterates over their representation\footnote{An implementation can be obtained upon request from the author}. We can easily parallelize the algorithm over the first loop; different parts of the database can be processed by different processors, kernels, or computers. 
\paragraph{Algorithm to determine NN} 
\begin{tabbing}
\textbf{forall} \= $\{\textbf{x}^{(i)} \in DB | i \in \{1.. s\}\}$\\ 
\> \{  \= \\
\>\>$for$\= $(k=t;k \neq 0, k--)$ \\
\>\>\>\{ \=  \\
\>\> \>\> $load(U_k(\textbf{x})^{(i)})$;\\
\>\>\>\>/* 1-Lipschitz property */\\
\>\>\>\>$if$ \= $  (\| U_k(\textbf{x})^{(i)} - U_k(\textbf{y})) \|_p  >= \epsilon$)    \\
\>\>\>\>\>$break:$;\\
 \>\>\>\>\>$if$ \= ($k=0$) $print$ $\textbf{x}^{(i)}$ $is$ $NN$ of  $\textbf{y}$\\
\>\> \>\} \\
\>\} \\
\end{tabbing}
Each call of the 1-Lipschitz property costs $dim(U_k)$. The cost according to Equation \ref{eq:lsubcost} correspond to the number of 1-Lipschitz property calls, corresponding to the value $\sigma_k$. 

\subsection{960-dimensional vector space}

We apply computational procedure on high-dimensional data set of $100 000$ vectors of dimensionality $960$.  The vectors represent the GIST global descriptor of an image and are composed by concatenated orientation image histograms \cite{Jegou11}.  
The vector $\textbf{ x}$ of dimensionality $960$ is split into $480$ distinct sub-vectors of dimensionality $2$. The data points are described by $480$ covariance matrices $C_i$ for each subspace. For all points, the covariance matrices are computed iteratively. 
\begin{equation}
 \textbf{ x} =\underbrace{x_1,x_2}_{ C_1},\underbrace{x_3,x_4}_{ C_2}, \cdots \cdots,\underbrace{x_{479},x_{959}}_{C_{480}}.
\end{equation}
 The resulting $480$ projections, $\textbf{z}_i \cdot \textbf{z}_i^\top $, define the adaptive projection $A: R^{960} \mapsto R^{480}$. We apply the adaptive projection and the determination of the adaptive projection recursively.  The resulting family of projections,
\begin{equation}
 A_1: U_0 \mapsto U_1; A_2: U_1 \mapsto U_2;  \ldots; A_7: U_{6} \mapsto U_7 
 \end{equation}
defines the dimensionalities of the subspaces.  
\[ dim(U_0)=960 > dim(U_1)=480 >  dim(U_2)=240 > dim(U_3)=120  \]
\[ > dim(U_4)=60 > dim(U_5)=30  >  dim(U_6)=10  > dim(U_7)=5. \]
In Table \ref{tab_res1}, we indicate the mean costs according to Equation \ref{eq:lsubcost} using the $l_2$ norm. 
\begin{table} [h]
\begin{center}
\begin{tabular}  {|c|c|c|c|c|} \hline
projection &   $\epsilon$ for $\approx 52$ NN  & mean cost  & ratio\\ \hline \hline
orthogonal &  $6300$  &  $4584277$ & $21.38$ \\ \hline
adaptive & $6300$  & $4393127$ & $22.31$  \\ \hline
\end{tabular}
\caption{Mean ratio of list matching to the mean computation costs according to Equation \ref{eq:lsubcost}.   The values were determined over a disjunct sample of $S \subseteq DB$ with size $|S|=400$. The adaptive projection gives only a slight improvement. The diversion from the $m$-secting line according to Equation \ref{eq:abstand} is always $\ll 0.0001$. }  \label{tab_res1}
\end{center}
\end{table}

\subsection{12288-dimensional vector space}

The 12288-dimensional vector space corresponds to an image database that consists of $9.876$ 3-band RGB (Red, Green, Blue) images of size $128 \times 96$. Each color is represented by 8 bits \cite{Wichert08}.  Each of the tree bands of size $128 \times 96$ is tiled with rectangular windows $W$ of size $4 \times 4$.  The data points are described by $32 \times 24$ covariance matrices $C_i$ for each subspace, for each band.  The resulting $768=32 \times 24$ projections $\textbf{z}_i \cdot \textbf{z}_i^\top $ define the adaptive projection $A: R^{12288} \mapsto R^{768}$ for each band (Red, Green, Blue).   We apply the adaptive projection and the determination of the adaptive projection recursively.  The resulting family of projections, 
 \begin{equation}
  A_1: U_0 \mapsto U_1; A_2: U_1 \mapsto U_2; A_3: U_{2} \mapsto U_3 
\end{equation}
defines the dimensionalities of the subspaces for each band  
\[ dim(U_0)=12288 > dim(U_1)=768 >  dim(U_2)=48=8 \times 6 > dim(U_3)=12=4 \times 3.  \]
For an orthogonal projection, the sequence of subspaces $U_0 \supset U_1 \supset U_2 \supset U_3$  corresponds to the ``image pyramid'' \cite{Burt83}, \cite{Gonzales01}, which has a base that contains an image with a high-resolution and an apex that is the low-resolution approximation of the image.  In Table \ref{tab_res2}, we indicate the mean costs according to Equation \ref{eq:lsubcost}. The $l_1$ norm gives the best results.
\begin{table} [h]

\begin{center}
\begin{tabular}  {|c|c|c|c|c|c|} \hline
projection & $l_p$ & $\epsilon$ for $\approx 52$ NN & cost  & ratio\\ \hline \hline
orthogonal & $l_1$ & $1240000$ & $8571752$ &  $42.47$ \\ \hline
adaptive & $l_2$ & $8500$  & $10343766$ & $35.20$  \\ \hline
orthogonal & $l_2$ & $8500$  & $10386043$ & $35.05$\\ \hline
orthogonal & $l_4$ & $825$  &  $12464281$ & $29.32$ \\ \hline
orthogonal & $l_{\infty}$ & $161$  &  $39639239$ & $9.19$ \\ \hline
\end{tabular}
\caption{Mean ratio of list matching to the mean computation costs according to Equation \ref{eq:lsubcost}.   The values were determined over a disjunct sample of $S \subseteq DB$ with size $|S|=400$. The diversion from the $m$-secting line, according to Equation  \ref{eq:abstand}, is always $\ll 0.0001$. }  
\label{tab_res2}
\end{center}
\end{table}

\section{Conclusion}

An adaptive projection that satisfies the 1-Lipschitz property defined by the first principal component was introduced.  We indicated the behavior of the projections for the $l_p$ norms. The Manhattan distance $l_1$ loses the least information, followed by Euclidean distance function. Most information is lost when using the Chebyshev distance function.
Motivated by the tree structure, we indicated a family of projections that defines a mapping that satisfies the 1-Lipschitz property.  It is composed of orthogonal or adaptive projections in the $l_p$ space. Each projection is applied recursively in a low-dimensional space, where ``the curse of dimensionality'' conjecture does not apply. 

\section*{\uppercase{Acknowledgements}}

The author would thank  \^Angelo Cardoso for the valuable suggestions. This work was supported by Funda\c{c}\~{a}o para a Ci\^{e}ncia e Tecnologia (FCT):  PTDC/EIA-CCO/119722/2010 and by Funda\c{c}\~{a}o para a Ci\^{e}ncia e Tecnologia FCT (INESC-ID multiannual funding) through the PIDDAC Program funds. 

\bibliographystyle{plain}

\end{document}